\documentclass[reprint,twocolumn,amsmath,amssymb,aps,prb]{revtex4}

\usepackage{graphicx}% Include figure files
\usepackage{dcolumn}% Align table columns on decimal point
\usepackage{bm}% bold math

 \usepackage{color}

\usepackage{amssymb}
\usepackage{graphicx}
\usepackage{amsbsy}

\newcommand\beq{\begin{equation}}
\newcommand\eeq{\end{equation}}
\newcommand\beqa{\begin{eqnarray}}
\newcommand\eeqa{\end{eqnarray}}
\newcommand{\nn}{\nonumber\\}

\newcommand{\NN}{s}

\newcommand{\dd}{\text{d}}
\newcommand{\ex}{\text{ex}}

\newcommand{\pure}{\text{pure}}

\begin{document}

%\preprint{}


\title{Note: An exact scaling relation for truncatable free energies of polydisperse hard-sphere mixtures}


\author{Andr\'es Santos}

\email{andres@unex.es}
\homepage{http://www.unex.es/eweb/fisteor/andres/}

\affiliation{Departamento de F\'{\i}sica, Universidad de
Extremadura, Badajoz, E-06071, Spain}

\date{\today}

\begin{abstract}
A theoretical model for polydisperse systems of hard spheres is said to be truncatable  when the excess free energy depends on the size distribution through a finite number $K$ of moments. This Note proves an exact scaling relation for truncatable free energies, which allows to reduce the effective degrees of freedom to $K-2$ independent combinations of the moments.
\end{abstract}

\date{\today}

%\keywords{}

\maketitle

Let us consider a polydisperse  hard-sphere  fluid mixture in $d$ dimensions with number density $\rho=N/V$, where $N$ is the total number of particles contained in a volume $V$. The number of particles with a diameter between $\sigma$ and $\sigma+\dd\sigma$ is $N x(\sigma)\dd\sigma$, so that $\int_0^\infty\dd\sigma \,x(\sigma)=1$. The  moments of the size distribution are
\beq
M_n\equiv \int_0^\infty\dd\sigma \,x(\sigma)\sigma^n.
\label{mom}
\eeq
The particular case of a discrete $\NN$-component mixture is included as $x(\sigma)=\sum_{i=1}^\NN x_i\delta(\sigma-\sigma_i)$ with $\sum_{i=1}^\NN x_i=1$.

Let $a^\ex[\rho,x(\sigma)]$  be the excess free energy per particle of the system in units of $k_BT$, where $k_B$ is the Boltzmann constant and $T$ is the absolute temperature. It is given by
\beq
a^\ex[\rho,x(\sigma)]=-N^{-1}\ln Q_N(V),
\label{aex}
\eeq
where
\beq
Q_N(V)=V^{-N}\int \dd\mathbf{r}_1\ldots\int \dd\mathbf{r}_N\prod_{1\leq i<j\leq N}\Theta\left(r_{ij}-\frac{\sigma_i+\sigma_j}{2}\right)
\label{QN}
\eeq
is the configuration integral, $\Theta(x)$ being the Heaviside step function.

We now  suppose that the mixture consists of a finite number $N_0=x_0 N$ of point particles (i.e., $\sigma_i=0$ for $1\leq i\leq N_0$) plus $N'=N-N_0$ particles ($N_0+1\leq j\leq N$) with a certain (continuous or discrete) size distribution $x'(\sigma)$. Thus,
\beq
x(\sigma)=x_0\delta(\sigma)+(1-x_0)x'(\sigma).
\label{x0}
\eeq
The number density of the ``bare'' mixture is $\rho'=N'/V=(1-x_0)\rho$ and its size moments  are related to those of the original composite mixture by
\beq
M_n'=\int_0^\infty\dd\sigma \,x'(\sigma)\sigma^n=\frac{M_n}{1-x_0},\quad n\geq 1.
\label{Mn}
\eeq
Setting $\sigma_i=0$  ($1\leq i\leq N_0$) in Eq.\ \eqref{QN} we get
\beq
Q_N(V)=(1-V'/V)^{N_0}Q_{N'}(V),
\label{QN'}
\eeq
where $V'=v_dN'M_d'=v_d N M_d$ is the volume excluded by the  $N'$ normal particles, $v_{d}=(\pi /4)^{d/2}/\Gamma (1+d/2)$ being the volume of a
$d$-dimensional sphere of unit diameter, and $Q_{N'}(V)$ is the configuration integral of the bare mixture.
In the derivation of Eq.\ \eqref{QN'} one first integrates over the
positions of the point particles, obtaining $(V-V')^{N'}$; the remaining integration over the normal
particles is then the same as in the system without point particles.
Therefore, in the special case \eqref{x0}, Eq.\ \eqref{aex} becomes
\beq
a^\ex[\rho,x(\sigma)]=-{x_0}\ln\left(1-v_d\rho M_d\right)+(1-x_0)a^\ex[\rho',x'(\sigma)].
\label{11b}
\eeq

Let us now go back to the general polydisperse case and assume a model free energy with ``truncatable'' structure,\cite{GKM82,SWC01,S02} i.e., the excess free energy $a^\ex[\rho,x(\sigma)]$ depends on the size distribution $x(\sigma)$ only through  a \emph{finite} number $K$ of moments $\{M_1, M_2,\ldots, M_K\}$. Dimensional analysis requires that the dependence  of $a^{\text{ex}}$ on $\rho$ and $\{M_1, M_2,\ldots, M_K\}$ takes place through the dimensionless combinations $\eta$ and $\{m_2,\ldots, m_K\}$, where $\eta\equiv v_d\rho M_d$ is the packing fraction and $m_n\equiv M_n/M_1^n$ are rescaled moments. Therefore, the truncatability hypothesis can be written as
\beq
a^\ex[\rho,x(\sigma)]=a^\ex(\eta;m_2,\ldots,m_K).
\label{1}
\eeq
It is important to bear in mind that Eq.\ \eqref{1} is not a rigorous property. For instance, Blaak\cite{B98} has exactly evaluated the fourth virial coefficient $B_4(\sigma_i,\sigma_j,\sigma_k,\sigma_\ell)$ for $d=3$ when the four diameters are such that the smallest sphere (say $\ell$) fits in the inner space made by the other three spheres being tangent to each other. The analytic expression of $B_4(\sigma_i,\sigma_j,\sigma_k,\sigma_\ell)$ is given as a linear combination of terms of the form $\sigma_i^{q_1}\sigma_j^{q_2}\sigma_k^{q_3}\sigma_\ell^{q_4}$ with $q_1+q_2+q_3+q_4=9$. However, while $q_1$, $q_2$, and $q_3$ are always not larger than $3$, terms with up to $q_4=9$ are present. This shows that the exact $B_4$ is, in contrast to what claimed in other works,\cite{B99b} incompatible with Eq.\ \eqref{1}, unless $K\geq 9$. Notwithstanding this, the truncatability hypothesis \eqref{1} is crucial from a practical point of view to reduce from functional to algebraic the phase transition problem  in polydisperse systems.\cite{SWC01} Moreover,
this hypothesis (with $K=3$) has recently received numerical support from simulation data of three-dimensional polydisperse mixtures,\cite{OL12} even for metastable states.

The objective now is to prove that, once the ansatz \eqref{1} is assumed, the exact relationship \eqref{11b} imposes a constraint under the form of a scaling law for $a^\ex(\eta;m_2,\ldots,m_K)$. First, note that, in the case of Eq.\ \eqref{x0}, Eq.\ \eqref{Mn} gives
\beq
m_n'=(1-x_0)^{n-1}{m_n},\quad n\geq 2.
\label{12}
\eeq
Next, assuming Eq.\ \eqref{1} and making use of Eq.\ \eqref{11b} one obtains
\begin{widetext}
  \beq
a^\ex(\eta;m_2,m_3,\ldots,m_K)+\ln(1-\eta)=\lambda\left[a^\ex(\eta;\lambda m_2,\lambda^2m_3,\ldots,\lambda^{K-1}m_K)+\ln(1-\eta)\right],
\label{scal}
\eeq
where $\lambda= 1-x_0$. The \emph{scaling} property \eqref{scal} is the main result of this Note. It implies the equivalent form
\beq
a^\ex(\eta;m_2,m_3,\ldots,m_K)
=
-\ln(1-\eta)+\frac{1}{m_2}
\mathcal{A}\left(\eta; \frac{m_3}{m_2^2},\ldots,\frac{m_K}{m_2^{K-1}}\right),
\label{A}
\eeq
\end{widetext}
where the scaling function $\mathcal{A}$ remains undetermined.
In the special case of a one-component system, $m_n=1$ and thus
\beq
\mathcal{A}(\eta;1,\ldots,1)=a^\ex_\pure(\eta)+\ln(1-\eta),
\label{pure}
\eeq
where $a^\ex_\pure(\eta)$ is the excess free energy  of the pure fluid. Scaling relations similar to \eqref{scal} and \eqref{A} can be obtained for the compressibility factor  $Z\equiv p/\rho k_BT$, where $p$ is the pressure,   by the thermodynamic relation $Z=1+{\eta}{\partial a^\ex}/{\partial\eta}$.

Equation  \eqref{A} constrains possible theoretical models with truncatable structure. In fact, it represents a significant reduction in the number of ``degrees of freedom'' of the excess free energy of a polydisperse HS fluid. The exact free energy $a^\ex$ is a \emph{functional} of the size distribution $x(\sigma)$ and thus it has an infinite number of degrees of freedom. The truncatability hypothesis reduces the number of independent variables to the first $K$ moments  $\{M_n, n=1,\ldots,K\}$, apart from the number density $\rho$. Dimensional analysis  trivially reduces that number to $K-1$ dimensionless moments $\{m_n, n=2,\ldots,K\}$, as indicated in Eq.\ \eqref{1}. Further, Eq.\ \eqref{A} makes explicit the dependence on $m_2$, thus reducing the effective number of independent quantities to $K-2$ ratios $m_n/m_2^{n-1}$ ($n=3,\ldots,K$).

The exact excess free energy in the one-dimensional case ($d=1$) is $a^\ex=-\ln(1-\eta)$ and thus it trivially verifies Eq.\ \eqref{A} with $\mathcal{A}=0$. The exact result for hard disks ($d=2$) is not known. However, assuming truncatability  with $K=2$, Eqs.\ \eqref{A} and \eqref{pure} imply
\beq
a^\ex(\eta;m_2)=-\ln(1-\eta)+\frac{1}{m_2}\left[a_\pure^\ex(\eta)+\ln(1-\eta)\right].
\label{K2}
\eeq
This result agrees with that derived by independent methods\cite{SYH99,SYH02,HYS08} and includes Jenkins and Mancini's equation of state\cite{JM87} as a particular case.
Finally, let us consider the three-dimensional case ($d=3$) with $K=3$. Now the scaling function $\mathcal{A}(\eta;y)$ depends on the ratio $y=m_3/m_2^2$ but otherwise it is arbitrary. Let us assume the explicit functional form $\mathcal{A}(\eta;y)=y^{-2}\left[\mathcal{A}_0(\eta)+\mathcal{A}_1(\eta)y\right]$.   Equation \eqref{pure} then implies that $\mathcal{A}_0(\eta)=a^\ex_\pure(\eta)+\ln(1-\eta)-\mathcal{A}_1(\eta)$. Therefore, Eq.\ \eqref{A} yields
\beqa
a^\ex(\eta;m_2,m_3)&=&-\ln(1-\eta)+\frac{m_2^3}{m_3^2}\left[a_\pure^\ex(\eta)+\ln(1-\eta)\right]\nn
&&+
\left(\frac{m_2}{m_3}-\frac{m_2^3}{m_3^2}\right)\mathcal{A}_1(\eta),
\label{K3}
\eeqa
where the function $\mathcal{A}_1(\eta)$ is not constrained by Eq.\ \eqref{pure}. Equation \eqref{K3} with $\mathcal{A}_1(\eta)=3\eta/(1-\eta)$ has been derived by a different route\cite{SYH05,HYS06,HYS08} and includes, as particular cases, the Scaled Particle
Theory,\cite{RFL59,HFL61,LHP65,MR75,R88,HC04b} the virial Percus--Yevick,\cite{L64} and the Boubl\'{\i}k--Mansoori--Carnahan--Starling--Leland\cite{B70,MCSL71} equations of state.
More generally, Eq.\ \eqref{K3} with free $\mathcal{A}_1(\eta)$ or Eq.\ \eqref{A} with $K=3$ and free $\mathcal{A}(\eta,y)$ can be useful to infer the equation of state of the pure system in the metastable region from measurements made on multi-component systems.\cite{SYH11}

The author is grateful to two anonymous reviewers for their constructive comments. Financial support from the Spanish Government through Grant No. FIS2010-16587 and from the Junta de Extremadura (Spain) through Grant No.\ GR10158 (partially financed by FEDER funds) is  acknowledged.

\bibliographystyle{apsrev}
\bibliography{D:/Dropbox/Public/bib_files/liquid}
\end{document}